\documentclass{PoS}

\usepackage{subcaption}

\title{Yellow symbiotic star AG~Draconis in the scope of the New Online Database of Symbiotic Variables}

\ShortTitle{AG~Dra in the New Online Database of Symbiotic Variables}

\author{\speaker{Jaroslav Merc}\\
        Faculty of Mathematics and Physics, Charles University, Ke Karlovu 3, 121 16 Prague 2, Czech Republic\\
        Faculty of Science, P. J. \v{S}af{\'a}rik University, Park Angelinum 9, 040 01 Ko\v{s}ice, Slovak Republic\\
        E-mail: \email{jaroslav.merc@student.upjs.sk}}

\author{Rudolf G{\'a}lis\\
        Faculty of Science, P. J. \v{S}af{\'a}rik University, Park Angelinum 9, 040 01 Ko\v{s}ice, Slovak Republic\\
        E-mail: \email{rudolf.galis@upjs.sk}}

\author{Laurits Leedj{\"a}rv\\
        Tartu Observatory, University of Tartu, Observatooriumi 1, T\~{o}ravere, 61602 Tartumaa, Estonia\\
        E-mail: \email{laurits.leedjarv@to.ee}}
        
\author{Marek Wolf\\
        Faculty of Mathematics and Physics, Charles University, Ke Karlovu 3, 121 16 Prague 2, Czech Republic\\
        E-mail: \email{marek.wolf@mff.cuni.cz }}

\abstract{Symbiotic stars are strongly interacting binaries, consisting of a white dwarf and a cool giant, mainly of spectral type M. AG~Draconis belongs to a less numerous group of the yellow symbiotic systems, as the cool component in this binary is of a spectral type earlier than K4. Recently, after seven years of quiescence, this symbiotic star exhibited a very unusual active stage with the four minor outbursts observed. Thanks to the excellent involvement of amateur astronomers and professional observatories, we can study the activity of AG~Draconis in unprecedented details. In the present work, we discuss the activity and peculiarities of this interacting system within the entire group of symbiotic stars whose properties have recently been presented in our New Online Database of Symbiotic Variables. }

\FullConference{The Golden Age of Cataclysmic Variables and Related Objects V (GOLDEN2019)\\
		2-7 September 2019\\
		Palermo, Italy}

\begin{document}

\section{AG~Draconis}
Symbiotic variables are the widest interacting binaries consisting of a cool giant (or a supergiant) of a spectral type K or M (rarely G) as a donor and a compact star, most commonly a hot white dwarf ($\approx 10^5$\,K), as the accretor of transferred matter (Miko\l ajewska, 2007). 

AG~Dra is one of the best studied symbiotic systems, but as discussed below, there are some unique features of this star. AG~Dra belongs to a less common group of the yellow symbiotic stars - the cool component is a red giant of an early spectral type (K0 - K4), with a low metallicity and a higher luminosity than that of a standard class III (giant). Its hot component is considered to be a white dwarf (WD) sustaining a high luminosity and temperature (Miko\l ajewska et al., 1995; Sion et al., 2012). The orbital period of this binary system is 551 days (Hric et al., 2014). 

The system undergoes characteristic symbiotic activity with alternating quiescent and active stages. The active stages occur in intervals of 9 - 15 years and consist of several outbursts repeating at about one-year interval (G\'{a}lis et al., 2017). The activity of AG~Dra usually begins with the major, prominent outburst followed by several minor ones.

As previously indicated by UV observations (Gonz\'{a}lez-Riestra et al., 1999), the characteristics of the optical emission lines clearly confirmed the presence of two types of the outbursts of AG~Dra: major ones are usually \textit{cool} and smaller-scale outbursts are of the \textit{hot} type (Leedj{\"a}rv et al., 2016). During the \textit{hot} outbursts, the brightness of AG~Dra is more or less linearly correlated with variations of emission lines characteristics (e.g. EWs) and the He\,{\sc ii} Zanstra temperature increases or remains unchanged. On the other hand, during the \textit{cool} outbursts, the He\,{\sc ii} Zanstra temperature drops as the pseudo-atmosphere of the WD expands.

\subsection{Recent outburst activity of AG~Dra}

\begin{figure}[p]
\centering
\begin{subfigure}{1\textwidth}
  \centering
  \includegraphics[width=\linewidth]{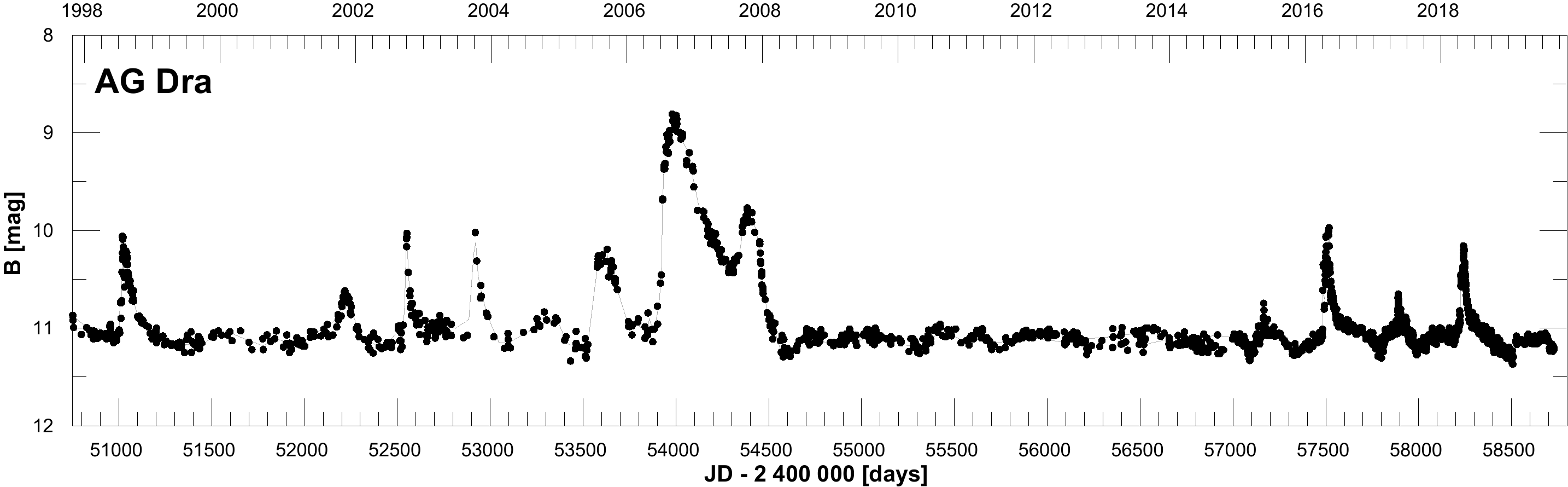}
\end{subfigure}\vspace{15pt}
\begin{subfigure}{1\textwidth}
  \centering
  \includegraphics[width=\linewidth]{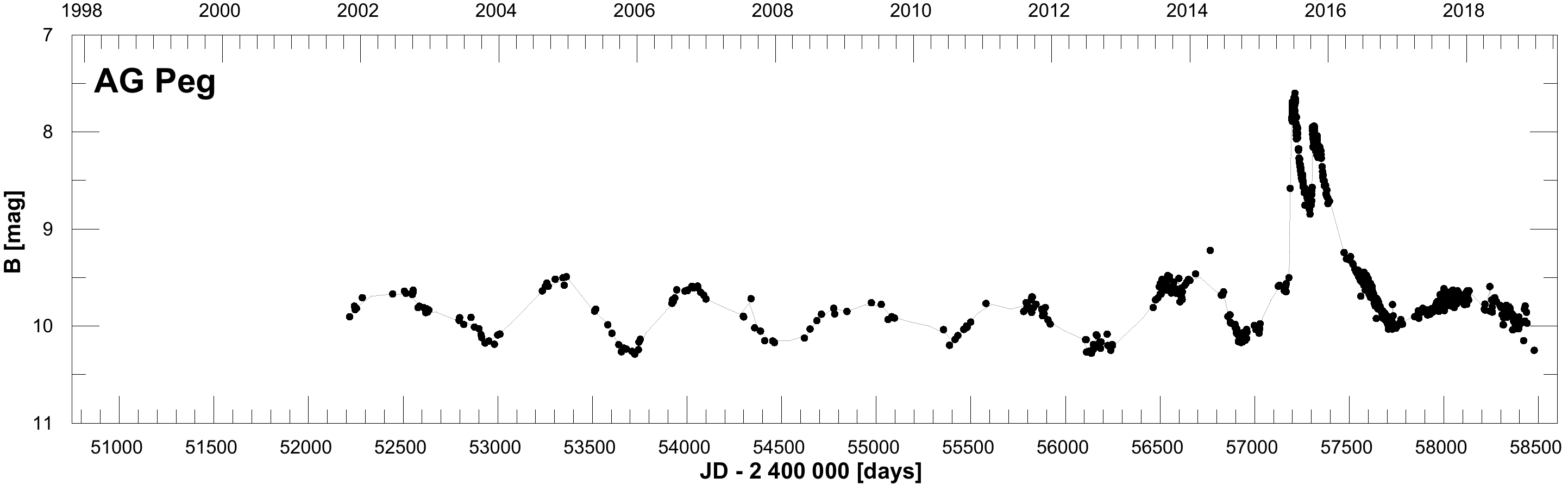}
\end{subfigure}\vspace{15pt}
\begin{subfigure}{1\textwidth}
  \centering
  \includegraphics[width=\linewidth]{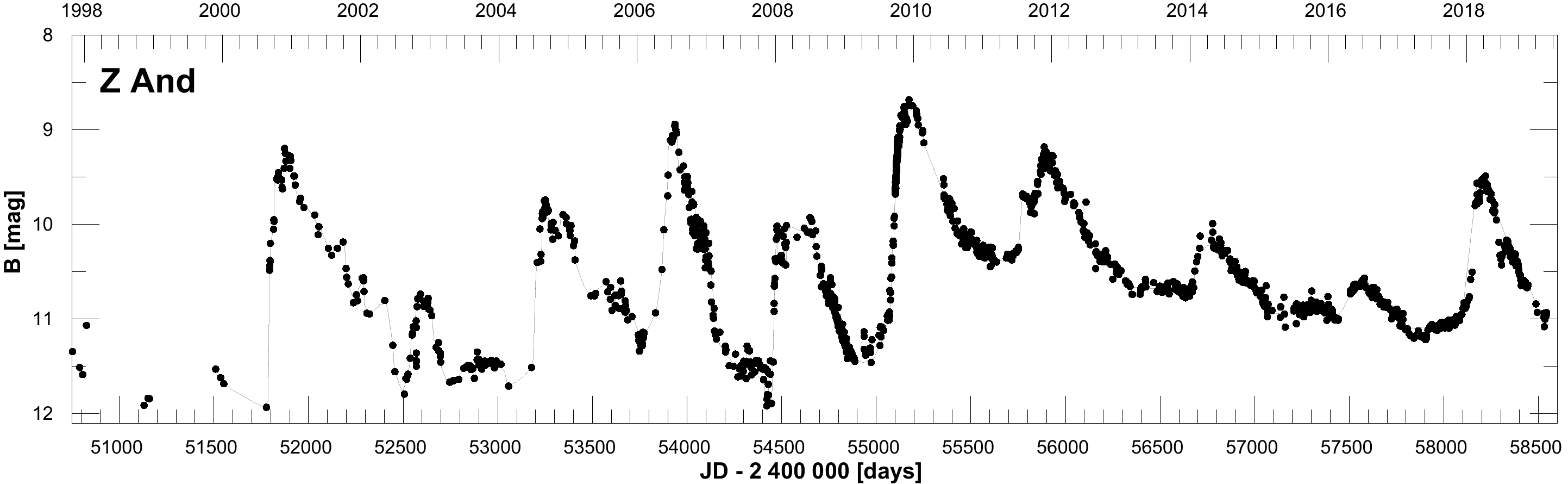}
\end{subfigure}\vspace{15pt}
\begin{subfigure}{1\textwidth}
  \centering
  \includegraphics[width=\linewidth]{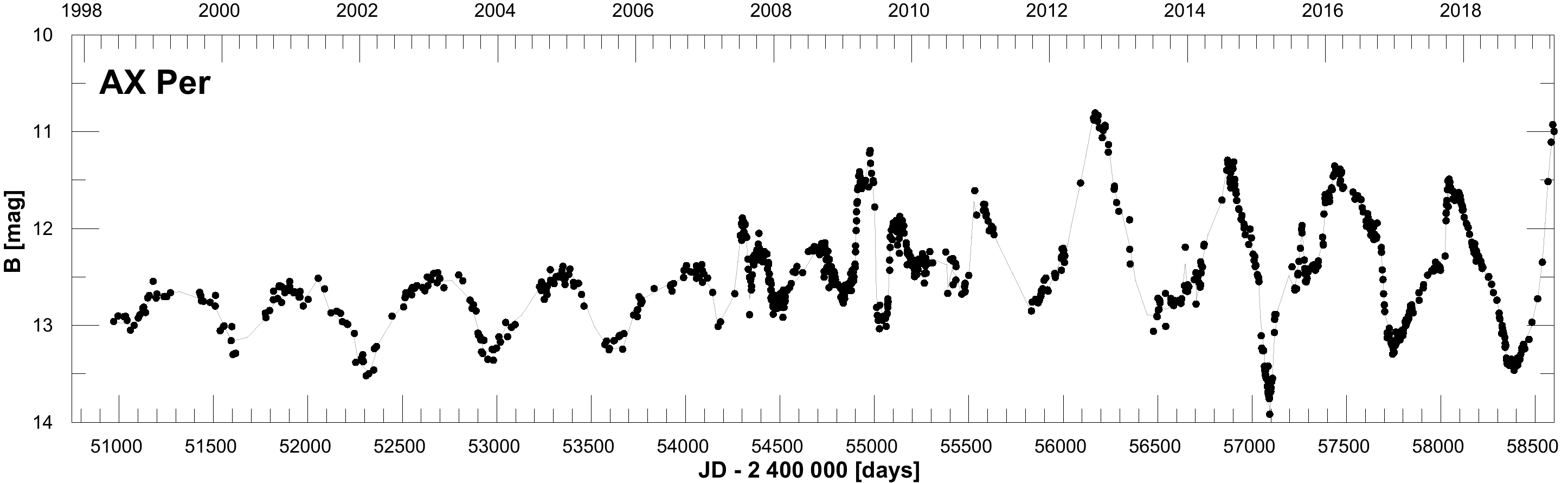}
\end{subfigure}
\caption{The light curves in $B$ filter over the period 1997 - 2019 of four galactic symbiotic stars which experienced outbursts in recent years. Data are obtained from AAVSO database (Kafka, 2019), Skopal et al. (2002, 2004, 2007, 2012) and Seker\' a\v s et al. (2019).}
\label{fig:comparisonLCs}
\end{figure}

Prodigious amount of observations of the symbiotic system have revealed that AG~Dra underwent at least six active stages since 1889. Previous, particularly long active stage (E+F) started in 1993 and continued until 2008 (Hric et al., 2014). This active stage was followed by seven years of quiescence (Q6) and AG~Dra begun rising again in brightness in the late spring of 2015 (see the top panel of Fig. \ref{fig:comparisonLCs}). We have observed series of four outbursts during the ongoing active stage (G\'{a}lis et al., 2019). 

According to the brightness of AG~Dra, all four recent outbursts can be classified as minor ones. The increase of EWs observed during these outbursts is indicating their \textit{hot} character (Merc et al., 2018). However, outbursts in the beginning of active phases of AG~Dra are usually major, \textit{cool} ones which makes the ongoing active stage very unusual. A detailed discussion of the features of this activity will be the subject of the forthcoming article.

\section{New Online Database of Symbiotic Variables}
AG~Dra is a variable star whose symbiotic character has been known for years (e.g. Kenyon, 1986). It was discovered accidentally like most symbiotic binaries in the previous century. However, systematic search for such objects has begun in recent decades and this effort has already brought a lot of new results. Surveys have led to the discovery of many new objects and dozens of candidates in the Milky Way (e.g. Miszalski et al., 2013; Miszalski \& Miko\l ajewska, 2014) and the Local Group (e.g. Gon\c calves et al., 2008, 2012, 2015; Kniazev et al., 2009; Miko\l ajewska et al., 2014, 2017; Shara et al., 2016; Roth et al. 2018). Subsequently, as the number of known systems grew, the catalog of symbiotic stars which was published by Belczy\'nski et al. (2000) became rapidly outdated. For this reason, we have prepared a new online database of the galactic and extragalactic symbiotic systems (Merc et al., 2019a).

The database is divided into two main parts according to the location of symbiotic variables. The first part consists of 74 confirmed and 88 suspected extragalactic symbiotic systems which are located in 14 galaxies (LMC, SMC, Draco Dwarf, IC\,10, M31, M33, M81, M87, NGC\,55, NGC\,185, NGC\,205, NGC\,300, NGC\,2403, NGC\,6822). The second part of the database consists of more than 480 galactic objects. The data of symbiotic variables are presented in the form of tables, which can be explored directly through the dedicated web-portal or can be downloaded and used offline in different formats (csv, xlsx, txt and pdf). Moreover, for all symbiotic binaries included in the database, we have prepared their object pages covering all available information, references, notes, and useful links. The database is accessible through the web-page: http://astronomy.science.upjs.sk/symbiotics/.

\subsection{AG~Dra in the New Online Database}
With the increasing number of known symbiotic stars, the vast amount of observations obtained particularly thanks to amateur observers and the availability of the New Online Database of Symbiotic Variables, symbiotic binaries can be studied as a whole population of interesting variable stars. In addition, individual stars can be studied in relation to other objects from the known population, allowing comparison of parameters and behavior. 

Several symbiotic stars have been studied very extensively during previous decades. Among them, AG~Dra is an exceptional case with hundreds of articles available on the Astrophysics Data System mentioning this object. However, there are still many open questions regarding the components of this interacting system or the outburst mechanisms as well. Moreover, it is not only the number of observations and studies, which is unique about the star.

\begin{figure}[t]
\centering
\includegraphics[width=0.95\linewidth]{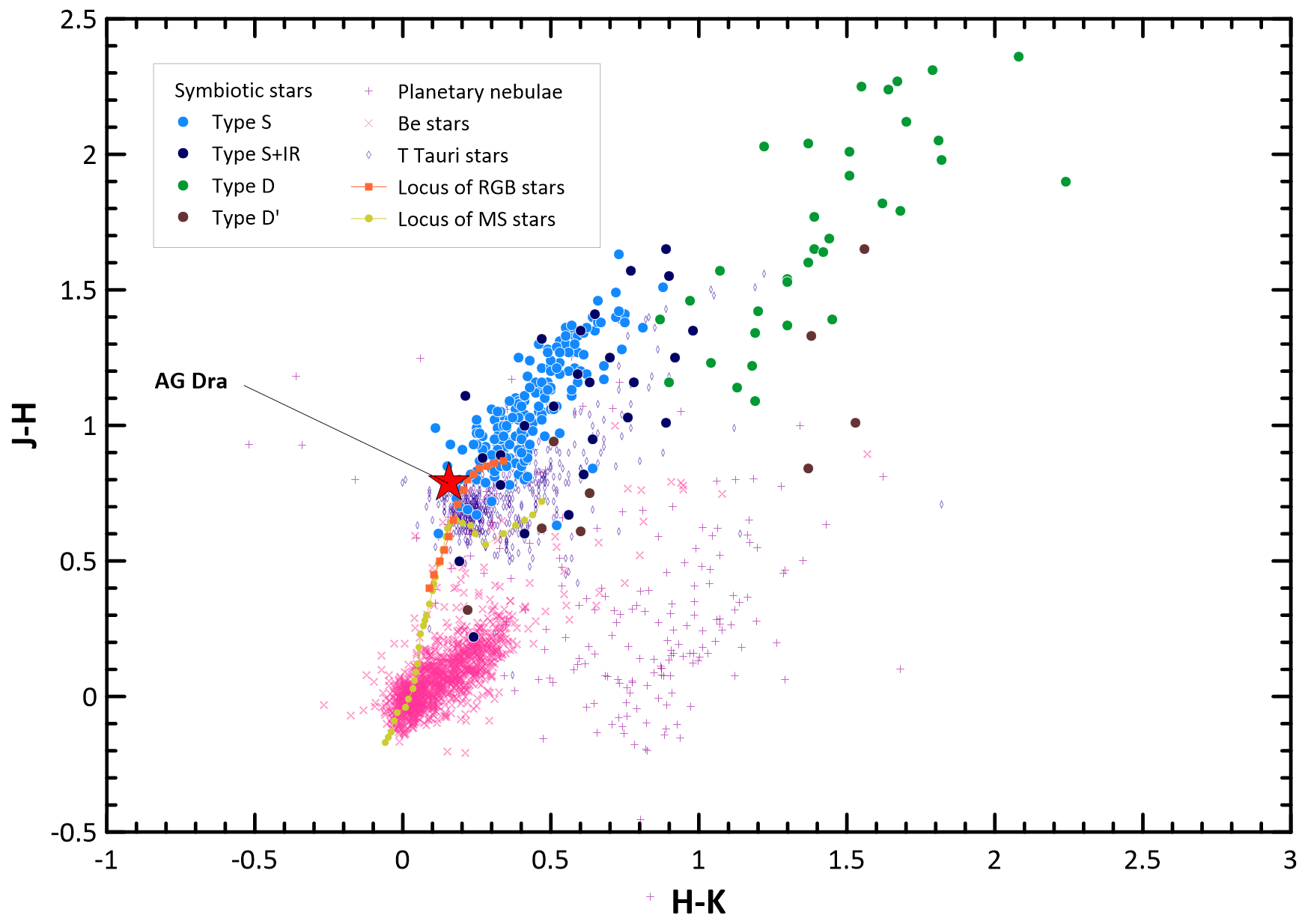}
\caption{The position of symbiotic stars of various infrared types (S, S+IR, D and D') as well as other objects in the near IR color-color diagram based on 2MASS observations. The Main sequence (MS) and Red giant branch (RGB) loci are taken from Strai\v zys \& Lazauskait\. e (2009), Be stars from Zhang et al. (2005), planetary nebulae from Ramos-Larios \& Phillips (2005) and T~Tauri stars from Dahm \& Simon (2005).}
\label{fig:ir_colors}
\end{figure}

AG~Dra is a classical symbiotic star of the infrared type S (Friedjung et al., 1998). Majority of the objects in the present version of the New Online Database of Symbiotic Variables ($\approx 83\,\%$) belongs to this type (Fig. \ref{fig:ir_colors}). The radius of the giant was estimated to be $33\pm11 \rm\, R_\odot$ (Skopal, 2005) and if we assume the volume radius of its corresponding Roche lobe to be $170 \rm \,R_\odot$ (Ogley et al., 2002), the giant is under-filling it, which is typical for most symbiotic binaries. 

\begin{figure}[t]
\centering
\includegraphics[width=0.95\linewidth]{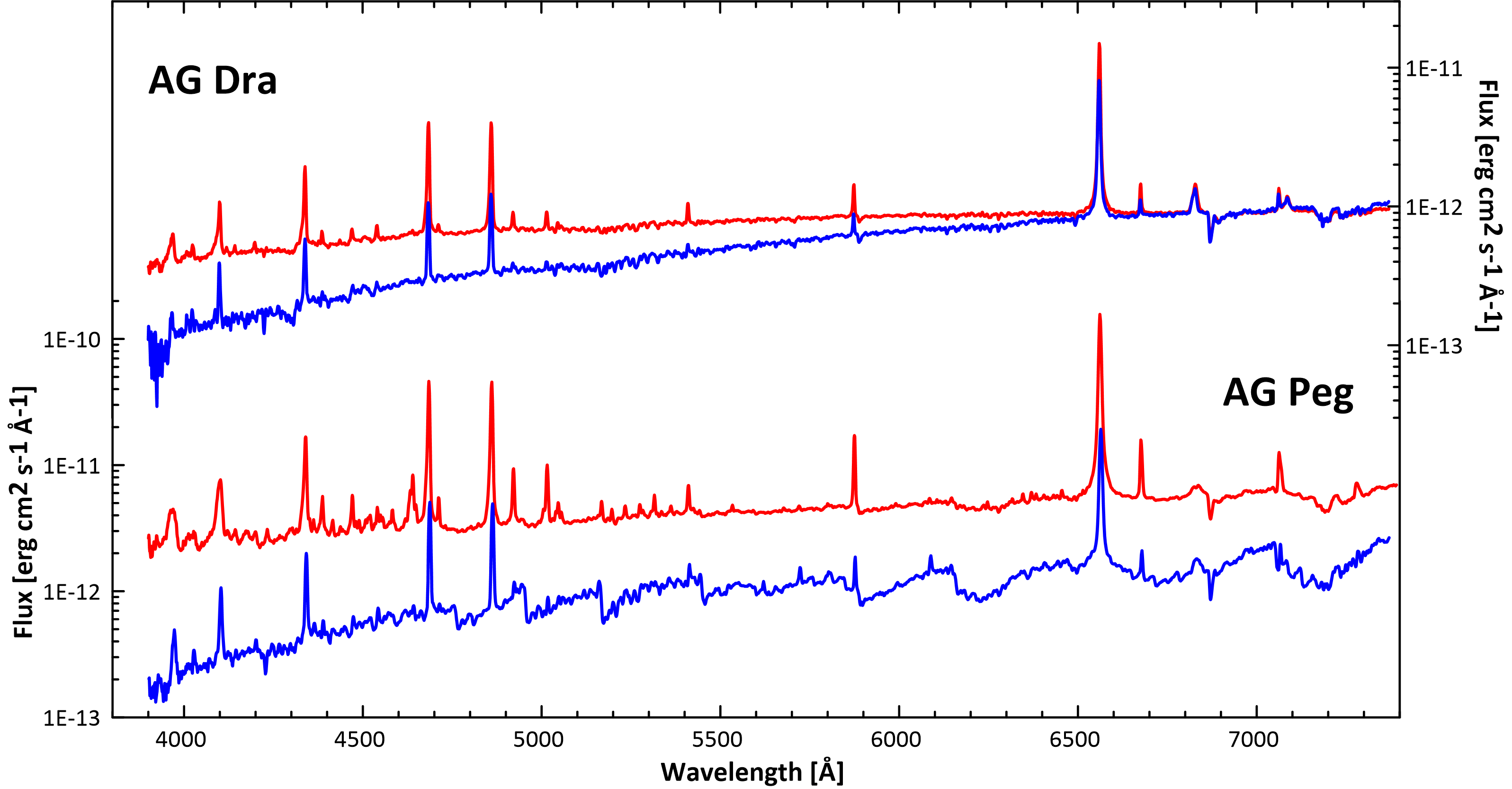}
\caption{Optical spectra of AG~Dra (the giant's spectral type K0 - K4) and AG~Peg (the giant's spectral type M3) during their quiescence (blue) and outburst (red) stages. Data are obtained from the ARAS database (Teyssier, 2019).}
\label{fig:spectrum}
\end{figure}

\begin{figure}[p]
    \centering
    \begin{subfigure}{1\textwidth}
  \centering
  \includegraphics[width=\linewidth]{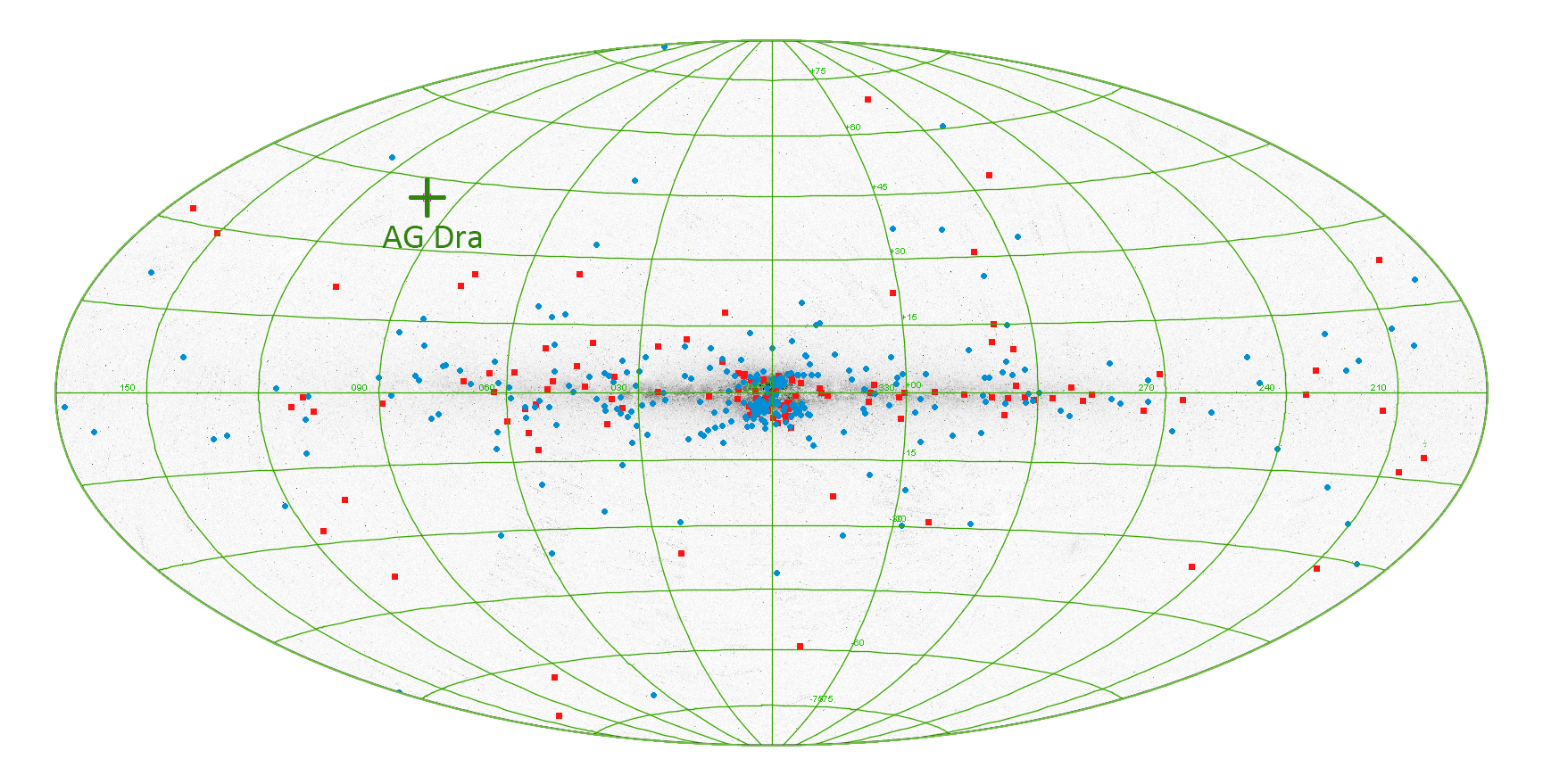}
  \caption{The distribution of galactic symbiotic stars overlaid on the 2MASS infrared image of the sky (in the galactic coordinates). Confirmed and suspected symbiotic stars are denoted by blue dots and red squares, respectively. The position of AG~Dra is shown by the green cross.}
  \label{fig:distribution}
\end{subfigure}\vspace{5pt}
    \begin{subfigure}{.465\textwidth}
  \centering
  \includegraphics[width=\linewidth]{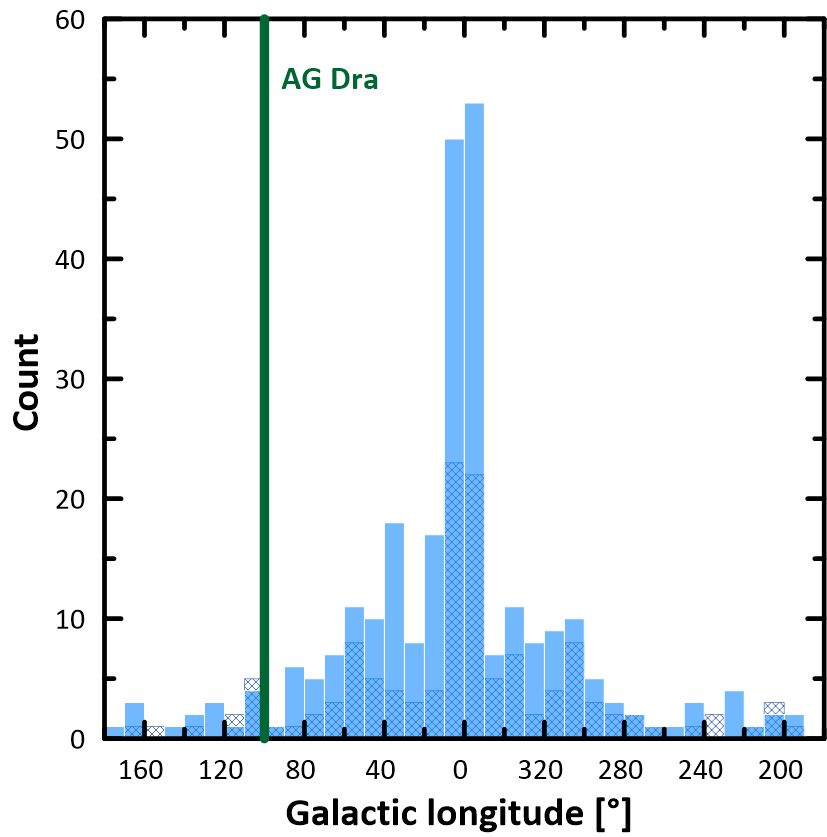}
  \caption{Histogram showing the distribution of symbiotic stars over the galactic longitude. The position of AG~Dra is shown by the green line.}
  \label{fig:histogram_x}
\end{subfigure}\hfill
\begin{subfigure}{.48\textwidth}
  \centering
  \vspace{4pt}
  \includegraphics[width=\linewidth]{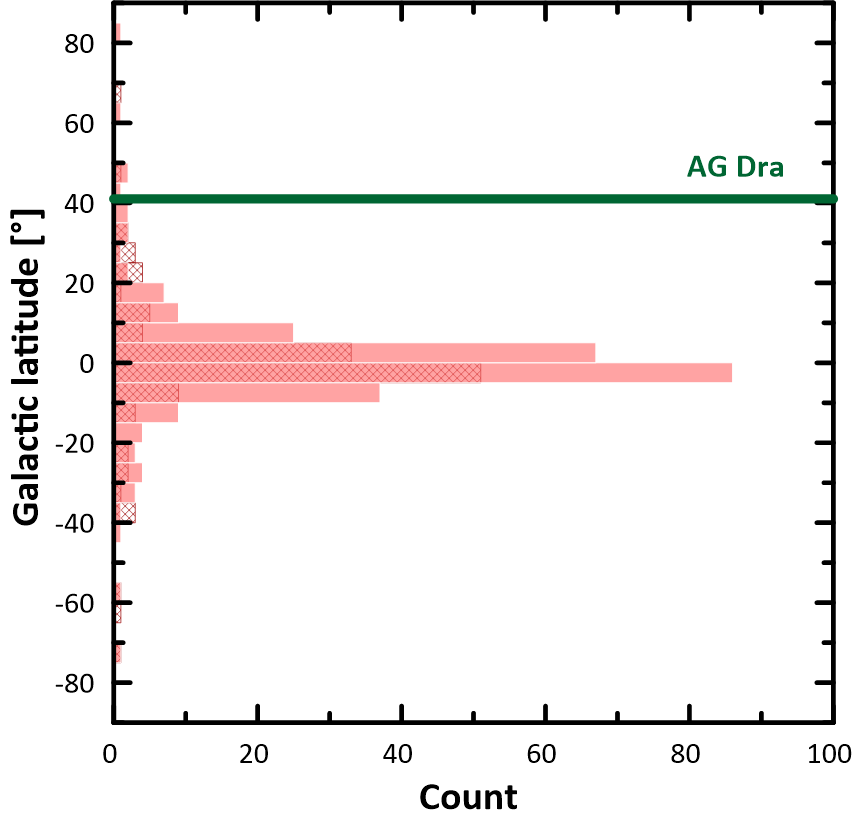}\vspace{3pt}
  \caption{Histogram showing the distribution of symbiotic stars over the galactic latitude. The position of AG~Dra is shown by the green line.}
  \label{fig:histogram_y}
\end{subfigure}
    \caption{The position of AG~Dra on the sky.}
    \label{fig:position}
\end{figure}

On the other hand, AG~Dra belongs to a small group ($\approx 10\,\%$) of so-called yellow symbiotic stars characterized by an early spectral type of the giant (K or even G) which is different from more typical spectral type M. While the exact spectral type of the giant in the AG~Dra system is not known, it is definitely of an early spectral type (K0 - K4; Fig. \ref{fig:spectrum}). The metallicity of the giant is very low (Fe/H $\approx -1.3$; Smith et al., 1996). Together with the radial velocity of the system of -147 km\,s$^{-1}$ and its position on the sky (Fig. \ref{fig:distribution}), these suggest that AG~Dra belongs to the old population of the galactic halo. Although the location of AG~Dra on the sky favors convenient long-term observations, at the same time it is not common for variable stars from this group, as nearly all galactic symbiotic stars are located around the Milky Way equator ($|b|<15^{\circ}$). For the objects in the current version of the database, 89\,\% of the confirmed and 83\,\% of candidate symbiotic stars are located in this sky region (Merc et al., 2020; Fig. \ref{fig:histogram_x} and \ref{fig:histogram_y}). 

The hot component of AG~Dra is a WD. Although some symbiotic binaries with a neutron star as an accretor have been discovered (e.g. Enoto et al., 2014), majority of them contain accreting WDs. Quasi-steady shell burning of the hydrogen-rich material on the surface of WD is probably also the source of super-soft X-ray emission of AG~Dra. There are 58 other symbiotic stars which have been detected in X-rays (Merc et al., 2019c). Nine sources, AG~Dra among them, show the super-soft emission which have been classified as the $\alpha$ type in the scheme introduced by Muerset et al. (1997) and extended by Luna et al. (2013). In addition to nine $\alpha$ sources, 18 have been classified as the $\beta$ type in which the X-ray emission is interpreted as due to the shock-heated plasma emerging in the collision of winds from the components, nine are neutron star accretors (the $\gamma$ type), and twelve have been classified as the $\delta$ type in which X-ray emission originates from the boundary layer between the accretion disk and the WD. Eleven systems are showing characteristics of both $\beta$ and $\delta$ types and are therefore classified as the  $\beta/\delta$ type. 

Very prominent emission lines are observed in optical spectra of the $\alpha$ and $\beta$ types (Akras et al., 2019; Fig. 2 in Merc et al., 2019c). In the case of AG~Dra, the emission lines of H\,{\sc i}, He\,{\sc i}, He\,{\sc ii} and Raman-scattered O\,{\sc vi} are observed almost all the time, with the emission lines of Fe\,{\sc ii} appearing during the outbursts and the O\,{\sc vi} lines vanishing during the \textit{cool} outbursts. The Raman-scattered O\,{\sc vi} lines, which are typically considered as an evidence of the symbiotic nature of an object are observed in $\approx 55 \,\%$ of symbiotic stars (Akras et al., 2019).

The spectral characteristics of symbiotic systems are changing dramatically throughout their outbursts (comparison of the quiescent and active optical spectra of AG~Dra is shown in Fig. \ref{fig:spectrum}). The AG~Dra binary star manifests the Z~And-type activity, characterized by recurrent outbursts with amplitudes of $\approx 1-3\,$mag at optical wavelengths that are repeating at intervals of about one or several years. The shape and duration of the outbursts of symbiotic stars showing this type of activity differ not only from object to object but also for one symbiotic binary during various activity stages. The AG~Dra is an excellent example of this variability as its light curve clearly demonstrates (Fig. \ref{fig:comparisonLCs}). 

In addition, several other mechanisms contribute to the complexity of the AG~Dra light curve. Although this symbiotic system is not an eclipsing one, the photometrically asymmetric nebula and its changing visibility during the orbital revolution are responsible for the sinusoidal variations of the light curve recognizable mainly during the quiescent stages. Due to the early spectral type of the giant, the photometric variations of AG~Dra are most significant at shorter wavelengths, particularly in $U$ filter, and are not so pronounced in $B$ or $V$ filters. In symbiotic stars comprising M giants, these variations are easily observable at least in $B$ filter (Fig. \ref{fig:comparisonLCs}), because the giant dominates at longer wavelengths. In case of AG~Dra, the light curves in the $B$ or $V$ filters are dominated by modulations with a period of around 355\,days, explained by the pulsations of the giant as suggested by G\'{a}lis et al. (1999).

Comparing of the light curves of outbursting symbiotic stars is interesting for several reasons. The similarities may indicate the same nature of the outburst mechanisms responsible for their observed activity. For example, Sokoloski et al. (2006) proposed the combination nova model as an explanation of the outburst activity of Z~And, in which smaller-scale \textit{hot} outbursts are explained
by the accretion disc instability model, as in the dwarf novae (Warner
1995) and major \textit{cool} outbursts are due to the thermonuclear runaway as in the classical nova outbursts. This model may also be applicable in the case of AG~Dra (Leedj{\"a}rv et al., 2016) and other symbiotic stars. 

On the other hand, it was very exciting to observe the symbiotic system AG~Peg to show the Z~And-type outburst in 2015 (Skopal et al., 2017, Merc et al., 2019b), 165 years after its nova-like flare-up. In the case of AG~Dra, very long period of quiescence had also been observed since the beginning of observations in 1889 until 1927 (or maybe even 1932). It is therefore possible, that AG~Dra (and other classical symbiotic stars as well) experienced the symbiotic nova stage in past and then started to show Z~And-type activity. Comparing the current behavior of these symbiotic systems can help to solve this issue. 

\section{Comparison of AG~Dra and LT~Del}

\begin{table}[]
\centering
\caption{Comparison of the parameters of AG~Dra and LT~Del.}
\label{tab:my-table}
\begin{tabular}{l|cc}\hline\hline
& \textbf{AG~Dra} & \textbf{LT Del} \\\hline
\textbf{Temperature of the giant {[}K{]}} & 4300 & 4400 \\
\textbf{Metallicity {[}Fe/H{]}} & -1.3 & -1.1 \\
\textbf{Radial velocity {[}km/s{]}} & -147 & -107 \\
\textbf{IR type} & S & S \\
\textbf{Activity} & Z And & Z And \\
\textbf{Orbital period {[}d{]}} & 551 & 476 \\
\textbf{Dominant spectral lines} & H I, He I, He II & H I, He I, He II \\
\textbf{Fe II lines} & During some outbursts & No \\
\textbf{O VI lines} & Yes & No \\
\textbf{X-rays} & Yes & No
\end{tabular}
\end{table}
As already mentioned, the cool component of AG~Dra is the giant of a spectral type K. Most of the symbiotic stars mentioned in the previous section (e.g. AG~Peg, Z~And) are comprising M giants. When looking for an object markedly similar to AG~Dra, besides an early spectral type of the giant (a yellow symbiotic star), it should meet several other criteria: low giant metallicity, infrared type S, Z~And-type activity, and it should be regularly observed to some extent, allowing comparison of both objects. The New Online Database of Symbiotic Variables can be used effectively for such a search.

One of the relatively appropriate candidates is the yellow, infrared type S symbiotic star LT~Del. The binary system with an orbital period of 476 days (Arkhipova et al., 2011) is consisting of a giant of a spectral type K3 with a low metallicity (Fe/H $\approx$ -1.1; Pereira et al., 1998) and a hot WD with a temperature of $\approx 10^5$\,K. The symbiotic system LT~Del, similarly to AG~Dra, belongs to the halo population which is suggested by the low metallicity and high radial velocity (-107 km\,s$^{-1}$). Comparison of the parameters of AG~Dra and LT~Del is listed in Tab. \ref{tab:my-table}.

During the observation period, LT~Del experienced two outbursts (in 1994 and 2017). The latter one has been studied by Ikonnikova et al. (2019). They reported that the outburst was of the \textit{hot} type, in contrast with the one observed in 1994 during which the temperature did not change compared to its quiescent values. As mentioned before, such behavior is typical for the symbiotic system AG~Dra. Similarly to AG~Dra, the quiescent light curves of LT~Del are dominated by sinusoidal variations caused by the changing visibility of emitting regions during the orbital revolution of the binary system. Ikonnikova et al. (2019) reported similar variability observed also in the case of the EWs of low excitation emission lines.

The main difference of the two symbiotic stars is in the presence of  Raman-scattered O\,{\sc vi} emission lines, which have not been detected in spectra of LT~Del. This symbiotic star has not been detected in the X-rays as well which points to a difference in the hot components of these systems or in geometry and optical thickness of the symbiotic nebulae.

\section{Conclusions}
Symbiotic stars are important astrophysical laboratories in studying a great variety of interesting phenomena. Yellow symbiotic stars are of particular interest because their absorption spectra are not influenced by TiO molecule absorption features which allows more precise measurements of atomic lines. The symbiotic system AG~Dra is the prototype of this small but growing group. Recent efforts have resulted in several new discoveries that allow for more systematic studies of the symbiotic population. 

The purpose of this work was not only to present the New Online Database of Symbiotic Variables but also to show how it can be utilized for the study of specific symbiotic system, in our case of AG~Dra. The binary aroused the interest recently when it showed very unusual activity stage and therefore it is more than ever interesting to compare the behavior of AG~Dra to other symbiotic stars, e.g. those showing Z~And-type activity, those containing K giants or those with the super-soft X-ray emission. As we have demonstrated, there are several options to select samples for comparison based on similarities or differences of AG~Dra to other known symbiotic stars. Based on this example, the New Online Database of Symbiotic Variables can be used for studies of other interesting symbiotic systems.

\section*{Acknowledgements}
This research was supported by the Slovak Research and Development Agency grant No. APVV-15-0458, by the Faculty of Science, P. J. \v{S}af\'{a}rik University in Ko\v{s}ice under the internal grant VVGS-PF-2019-1047, by the Charles University, project GA UK No. 890120, and by the Estonian Ministry of Education and Research institutional research funding IUT 40-1.

\bigskip
\bigskip
\noindent {\bf DISCUSSION}

\bigskip
\noindent {\bf ALESSANDRO EDEROCLITE:} Is the database VO-compatible?

\bigskip
\noindent {\bf JAROSLAV MERC:} Not in the present form. However, thank you for the comment, we may think about such feature in the future.

\end{document}